\documentclass[12pt]{article}

\usepackage{amsmath}

\renewcommand{\theequation}{\thesection.\theequation}
\numberwithin{equation}{section}

\newcommand{\N}{N\raise.7ex\hbox{\underline{$\circ $}}$\;$}

\begin{document}

\title{ E.M. Ovsiyuk\footnote{e.ovsiyuk@mail.ru},  O.V. Veko \\
Particle with spin 1 in a magnetic field \\on the hyperbolic
plane $H_{2}$ \\
{\small Mozyr State Pedagogical University named after I.P. Shamyakin, Belarus}}

\date{}

\maketitle

\begin{abstract}
There are constructed exact solutions of the quantum-mechanical
equation for a spin $S=1$ particle in 2-dimensional Riemannian
space of constant negative   curvature,
 hyperbolic   plane,  in presence of an external magnetic field,
 analogue of the homogeneous magnetic field in the Minkowski  space.
  A generalized formula for energy levels describing   quantization of the motion
  of the vector particle in magnetic field on the 2-dimensional space  $H_{2}$ has been found,
  nonrelativistic   and relativistic equations
  have been solved.

\end{abstract}

\subsection*{1.  Introduction}

The  quantization  of a quantum-mechanical particle in the
homogeneous magnetic field belongs to classical  problems in
physics  \cite{1,2,3, Landau-Lifshitz}. In 1985 -- 2010, a more
general problem in a curved Riemannian background,  hyperbolic and
spherical  planes,  was extensively studied \cite{Comtet-1985,
Comtet-1987,
 Aoki-1987, Groshe-1988,
Klauder-Onofri, Avron-Pnueli-1990, Plyushchay-1991(1),
Plyushchay-1991(2), Dunne-1992, Plyushchay-1995,
Alimohammadi-Shafei Deh Abad-1996, Alimohammadi-Mohseni
Sadjadi-1996, Onofri-2001, Negro et al-2001, Gamboa et al-2001,
Klishevich-Plyushchay-2002, Drukker et  al-2004,
Ghanmi-Intissar-2004, Correa-Jakubsky-Plyushchay-2009,
Alvarez-Cortes-Horvathy-Plyushchay-2009},
   providing us with a new system having intriguing
dynamics and symmetry, both on  classical and quantum levels.

Extension  to 3-dimensional hyperbolic and spherical spaces was
 performed recently.
In  \cite{4,5,6}, exact solutions for  a scalar particle in
extended problem, particle in  external magnetic field on the
background of Lobachevsky $H_{3}$  and Riemann $S_{3}$ spatial
geometries were found. A corresponding system in the frames of
classical mechanics was examined in
 \cite{7,8,9}. In the present paper, we consider a quantum-mechanical problem  a particle with spin $1/2$ described
 by the Dirac equation in  3-dimensional Lobachevsky  and Riemann space models
 in presence of the external magnetic field.

In the present paper, we will construct exact solutions for a
vector particle described by 10-dimensional  Duffin--Kemmer
equation in external magnetic field on the background of
2-dimensional spherical space  $H_{2}$.

10-dimensional Duffin--Kemmer equation for  a vector particle in a
curved space-time has the form \cite{Book-1}
$$
\left \{ \beta^{c} \left [ i  \; (\;  e_{(c)}^{\beta}
\partial_{\beta}  + {1\over 2} J^{ab} \gamma_{abc}  \; )\;
+\; {e \over \hbar c} A_{(c)}  \right  ]  -   {mc\over \hbar}
\right \} \Psi = 0 \; , \eqno(1.1)
$$

\noindent
where   $\gamma_{abc}$ stands for Ricci rotation
coefficiemts,
 $A_{a} = e_{(a)}^{\beta}  A_{\beta} $ represent tetrad components of electromagnetic 4-vector  $A_{\beta} $;
 $J^{ab} =  \beta ^{a}\beta ^{b}- \beta^{b} \beta ^{a}$ are generators of 10-dimensional
represen\-tation of the Lorentz group. For shortness, we use
notation $ e/c \hbar \Longrightarrow e, \; mc/ \hbar
\Longrightarrow M$.

In the space  $H_{3}$ we will use the system of cylindric coordinates \cite{Olevsky}
$$
dS^{2} =  c^{2}dt^{2} -  \mbox{ cosh}^{2} z ( d r^{2} + \mbox{sinh}^{2} r
\; d \phi^{2} ) - dz^{2}\; ,
$$$$
u_{1} = \mbox{cosh} \; z \; \mbox{sinh}\; r \cos \phi \; , \;\; u_{2}
= \mbox{cosh} \; z \; \mbox{sinh}\;r \sin \phi \; ,
$$$$
 u_{3}
= \mbox{sinh}\; z \; , \;\;  u_{0} = \mbox{cosh}\; z \; \mbox{cosh} \;
r \, ;
\eqno(1.2)
$$
$$
G = \{ \;r \in [ 0 , + \infty ) , \; \phi \in [ 0 , 2\pi ], \; z
\in (- \infty , + \infty ) \; \} \; .
$$

Generalized expression for electromagnetic potential for an homogeneous magnetic field in  the curved model $H_{3}$
is given as follows
$$
 A_{\phi} = -2B \sinh^{2} {r \over 2} = -B\; ( \cosh r -1 )\; .
\eqno(1.3)
$$

We will consider the above equation in presence of the field
in the model  $H_{3}$. Corresponding to cylindric coordinates  $x^{\alpha}= (t,r,\phi,z)$
a tetrad can be chosen as
$$
 e_{(a)}^{\beta}(x) = \left |
\begin{array}{llll}
1 & 0 & 0 & 0 \\
0 & \cosh^{-1}z & 0 & 0 \\
0 & 0 & \cosh^{-1}z\;\sinh^{-1} r & 0 \\
0 & 0 & 0 & 1
\end{array} \right |  .
\eqno(1.4)
$$

Taking into account relations
$$
\Gamma^{r}_{\;\;jk } = \left | \begin{array}{ccc}
0 & 0 & \tanh\;z \\
0 & - \sinh r \cosh r & 0 \\
\tanh\;z & 0 & 0
\end{array} \right | \; , \qquad
\Gamma^{\phi}_{\;\;jk } = \left | \begin{array}{ccc}
0 & \coth\,r & 0\\
\coth\,r & 0 & \tanh\, z \\
0 & \tanh\, z & 0
\end{array} \right | \; ,
$$
$$
\Gamma^{z}_{\;\;jk } = \left | \begin{array}{ccc}
-\sinh z \cosh z & 0 & 0\\
0 & -\sinh z \, \cosh z \,\sinh^{2} r & 0 \\
0 & 0 & 0
\end{array} \right | \; .
$$
$$
\gamma_{12 2} =
 { 1 \over \cosh z\, \tanh r} \; , \qquad
 \gamma_{31 1} =
 \tanh z\; , \qquad \gamma_{32 2} =
 \tanh z\; ,
 \eqno(1.5)
 $$

\noindent eq.  (1.1) reduces to
 the form
$$
\left \{  i \beta^{0} { \partial \over \partial t} + {1 \over \cosh z} \left (  i  \beta^{1}
{ \partial \over \partial r}        + \beta^{2}      { i  \partial _{ \phi} - e B
(\cosh r -1) + i J^{12} \cosh r  \over \sinh r}  \right ) +  \right.
$$
$$
 \left.
  +
   i\beta^{3}  { \partial \over \partial  z}  -  i   {\sinh z \over \cosh z}  \; ( \beta^{1} J^{13}    +   \beta^{2} J^{23} )
          - M   \right \} \Psi  = 0 \; .
\eqno(1.6)
$$

To separate the variables in eq. (1.5), we are to employ an explicit form
of the Duffin--Kemmer matrices $\beta^{a}$;  it will be most convenient to use so called cyclic representation
\cite{Book-5}, where  the generator  $J^{12}$ is of diagonal form (we specify matrices by blocks in accordance  with
( $1-3-3-3$)-splitting)
$$
\beta^{0} = \left | \begin{array}{rrrr}
 0       &   0        &  0  &  0 \\
 0  &  0       &  i  & 0  \\
  0  &   -i       &   0  & 0\\
   0  &  0       &   0  & 0
\end {array}
\right |, \qquad
\beta^{i} = \left |
\begin{array}{rrrrr}
  0       &  0       &    e_{i}  & 0       \\
    0   &  0       &   0      & \tau_{i} \\
   -e_{i}^{+}  &  0       &   0      & 0       \\
   0       &  -\tau_{i}&   0      & 0
\end {array} \right | \; ,
\eqno(1.7)
$$

\noindent where  $e_{i}, \; e_{i}^{\;t}, \; \tau_{i}$ denote
$$
e_{1} = {1 \over \sqrt{2}} ( -i, \; 0  , \; i )\; , \qquad e_{2} =
{1 \over \sqrt{2}} ( 1 , \; 0  , \;  1 )\; , \qquad e_{3} = ( 0 ,
i  , 0)\; , \;
$$
$$
\tau_{1} = {1 \over \sqrt{2}} \left |  \begin{array}{ccc} 0  &  1
&  0  \\ 1 &  0  &  1  \\ 0  &  1  &  0
\end{array} \right | , \qquad
\tau_{2}= {1 \over \sqrt{2}} \left |
\begin{array}{ccc} 0  &  -i  &  0  \\ i & 0  &  -i  \\ 0  &  i  &
0
\end{array} \right | , \qquad
  \tau_{3} =   \left |
\begin{array}{rrr} 1  &  0  &  0  \\ 0  &  0  &  0   \\ 0  &  0
&  -1
\end{array} \right |  =  s_{3}\; .
$$
$$
\eqno(1.8)
$$

\noindent  The generator  $J^{12}$ explicitly reads
$$
 J^{12} =  \beta^{1} \beta^{2} -  \beta^{2} \beta^{1}
=
$$
$$
= \left | \begin{array}{cccc}
(-e_{1}e_{2}^{+}  + e_{2}e_{1}^{+})& 0 & 0 & 0 \\
0 &(-\tau_{1}\tau_{2} + \tau_{2} \tau_{1}) & 0 & 0 \\
0 & 0 & (-e_{1}^{+} \bullet e_{2} + e_{2}^{+} \bullet e_{1}) & 0 \\
0 & 0 & 0 & (- \tau_{1}\tau_{2} + \tau_{2} \tau_{1})
\end{array} \right | =
$$
$$
=  -i \left | \begin{array}{cccc}
0 & 0  &  0 & 0  \\
0 &   \tau_{3}  & 0 & 0 \\
0 & 0 &  \tau_{3} & 0 \\
0 & 0 & 0 &  \tau_{3}
\end{array} \right | = -iS_{3}.
\eqno(1.9)
$$

\subsection*{2. Restriction to 2-dimensional model
 }

Let us restrict ourselves
to 2-dimensional case, spherical space  $H_{2}$ (formally it is sufficient in eq. (1.5)  to remove dependence on the variable
$z$  fixing its value by  $z=0$)
$$
\left [   i \beta^{0}  {\partial \over \partial t} + i   \beta^{1}
{\partial \over \partial r}        + \beta^{2}   \,   { i \partial
_{ \phi} - e B\, (\cosh r -1) + i J^{12} \cosh r  \over  \sinh r}
           -  M   \right ] \Psi  = 0 \; .
\eqno(2.1)
$$

 \noindent
With the use of substitution
$$
\Psi = e^{-i\epsilon t  }  e^{im\phi}   \left |
\begin{array}{c}
\Phi_{0}  (r) \\
\vec{\Phi}(r)  \\
\vec{E} (r)  \\
\vec{H} (r)
\end{array} \right | ,
\eqno(2.2)
$$

\noindent
eq.  (2.1)  assumes the form
 (introducing notation   $m + B ( \cosh r-1)  = \nu (r)$)
$$
\left [  \epsilon  \;  \beta^{0}   +  i   \beta^{1}\; {\partial
\over  \partial r }        - \beta^{2}    \;   { \nu (r)   -  \cosh
r \; S_{3}    \over  \sinh r}       -  M   \right ] \left |
\begin{array}{c}
\Phi_{0}  (r) \\
\vec{\Phi} (r) \\
\vec{E} (r) \\
\vec{H} (r)
\end{array} \right | =0 \;  .
\eqno(2.3)
$$

\noindent
Eq. (2.3)   reads
$$
 \left  [ \;  \epsilon \;  \left | \begin{array}{rrrr}
 0       &   0        &  0  &  0 \\
 0  &  0       &  i  & 0  \\
  0  &   -i       &   0  & 0\\
   0  &  0       &   0  & 0
\end{array} \right |    + i\;  \left |
\begin{array}{rrrrr}
  0       &  0       &    e_{1}  & 0       \\
    0   &  0       &   0      & \tau_{1} \\
   -e_{1}^{+}  &  0       &   0      & 0       \\
   0       &  -\tau_{1}&   0      & 0
\end {array} \right | {\partial \over \partial r } - \right.
$$
$$
\left.  - {1 \over  \sinh r  } \; \left |
\begin{array}{rrrrr}
  0       &  0       &    e_{2}  & 0       \\
    0   &  0       &   0      & \tau_{2} \\
   -e_{2}^{+}  &  0       &   0      & 0       \\
   0       &  -\tau_{2}&   0      & 0
\end {array} \right |
 (   \nu     - \cosh  r \; S_{3})
  - M\;   \right  ]  \left | \begin{array}{c}
\Phi_{0} \\
\vec{\Phi} \\
\vec{E} \\
\vec{H}
\end{array} \right |
 = 0 \;  ,
\eqno(2.4)
$$

\noindent or in a block form
$$
   i e_{1}  \partial _{r } \vec{E}
   -  {1 \over  \sinh r} \; e_{2} (\nu  - \cosh r \,s_{3} )  \vec{E}
     = M    \; \Phi_{0} \; ,
$$
$$
i \epsilon \cosh z\, \vec{E} + i \tau_{1} \partial_{r} \vec{H} -
{\tau_{2} \over \sinh r } ( \;  \nu  - \cosh r \, s_{3} \;) \vec{H}
=M  \; \vec{\Phi} \; ,
$$
$$
-i\epsilon \cosh z\, \vec{\Phi}  -i e_{1}^{+} \partial_{r} \Phi_{0}
+ {\nu \over  \sinh r} \,e_{2}^{+} \Phi_{0}  = M   \vec{E}\; ,
$$
$$
-i \tau_{1} \partial_{r} \vec{\Phi} + { (\nu -\cosh r \, s_{3})
\over  \sinh r} \,\tau_{2}  \vec{\Phi}  = M
\; \vec{H}\; .
\eqno(2.5)
$$

After separation of the variables we get
$$
  \gamma  (  { \partial  E_{1} \over \partial  r} - {\partial E_{3} \over \partial r
})  -  {\gamma \over  \sinh r }   \left [  (\nu -\cosh r )
E_{1} + (\nu +\cosh r) E_{3}   \right ]
 = M    \Phi_{0} \; ,
$$
$$
+i \epsilon \cosh z  E_{1}    +   i \gamma  {\partial H_{2} \over \partial  r}
 +   i\gamma     { \nu \over  \sinh r }   H_{2}
 = M  \Phi_{1}\;,
$$
$$
+i \epsilon   E_{2}  +  i \gamma ( {\partial  H_{1} \over \partial  r
} + {\partial  H_{3} \over \partial  r } ) -
 {i\gamma
\over  \sinh r } \left [  (\nu -\cosh r)  H_{1}  -   (\nu + \cosh r)  H_{3}  \right ]  =
M   \Phi_{2}\;,
$$
$$
+i \epsilon    E_{3}    +  i \gamma\; {\partial   H_{2} \over \partial
r}  -   i \gamma  {\nu  \over  \sinh r }  H_{2}  =
M   \Phi_{3}
$$
$$
\eqno(2.6)
$$

$$
-i  \epsilon   \Phi_{1}  +  \gamma   {\partial
\Phi_{0} \over d r}   +
 \gamma  {\nu \over  \sinh r }   \Phi_{0} = M    E_{1}\;,
$$
$$
-i \epsilon   \Phi_{2}   = M
E_{2}\;,
$$
$$
-i \epsilon \Phi_{3}  -   \gamma   {\partial  \Phi_{0} \over \partial
r}  +
 \gamma {\nu   \over  \sinh r }   \Phi_{0} = M  E_{3}\;,
$$
$$
\eqno(2.7)
$$

$$
-i  \gamma   {\partial  \Phi_{2} \over \partial  r}  -  i \gamma   {\nu   \over  \sinh r }  \Phi_{2} = M \cosh z  H_{1}\;,
$$
$$
 - i  \gamma  (  {\partial  \Phi_{1} \over \partial r } + { \partial \Phi_{3} \over \partial r })
+  {i \gamma \over  \sinh r }  [ (\nu -\cosh r) \Phi_{1} -
(\nu+\cosh r)\Phi_{3}  ]  = M   H_{2} \; ,
$$
$$
-i  \gamma   {\partial  \Phi_{2} \over \partial r}  +
 i \gamma {\nu \over  \sinh r }   \Phi_{2}
  =  M    H_{3}\;.
$$
$$
\eqno(2.8)
$$

With the notation
$$
{1 \over  \sqrt{2}}  ( {\partial  \over \partial r} +  { \nu -\cosh r  \over  \sinh r } ) = \hat{a}_{-} , \;
{1 \over  \sqrt{2}}  ( {\partial  \over \partial r} +  { \nu +\cosh r  \over  \sinh r } ) = \hat{a}_{+}\;, \;
{1 \over  \sqrt{2}}   ( {\partial  \over \partial r} +  { \nu   \over  \sinh r } ) = \hat{a} \; ,
$$
$$
{1 \over  \sqrt{2}}  (- {\partial  \over \partial r} +  { \nu -\cosh r  \over  \sinh r } ) = \hat{b}_{-} , \;
{1 \over  \sqrt{2}}  (- {\partial  \over \partial r} +  { \nu +\cosh r  \over  \sinh r } ) = \hat{b}_{+}\;, \;
{1 \over  \sqrt{2}}   (- {\partial  \over \partial r} +  { \nu   \over  \sinh r } ) = \hat{b} \; ,
$$

\noindent
the above system reads
$$
  -\hat{b}_{-}   E_{1}   - \hat{a}_{+}    E_{3}
 = M   \; \Phi_{0} \; ,
$$
$$
   - i  \hat{b}_{-}    H_{1}    +  i \hat{a}_{+} \; H_{3}
    + i \epsilon  \;   E_{2}   = M  \;  \Phi_{2}\;,
    $$
    $$
        i  \hat{a}     H_{2} +i \epsilon  \;   E_{1}
  = M  \;   \Phi _{1}\;,
$$
$$
  - i  \hat{b}  H_{2} +i \epsilon  \;   E_{3}
= M    \; \Phi_{3}\; ,
\eqno(2.9)
$$
$$
    \hat{a}  \Phi_{0}  - i  \epsilon  \;  \Phi_{1} = M \;     E_{1}\;,
$$
$$
-  i  \hat{a}  \Phi_{2}  = M  \; H_{1}\;,
$$
$$
   \hat{b}
\Phi_{0}  -i \epsilon  \;\Phi _{3}  = M    \; E_{3}\;,
$$
$$
  i   \hat{b}   \Phi_{2}
   =  M  \;    H_{3}\;,
$$
$$
-i \epsilon   \Phi_{2}    = M  \;  E_{2} \;,
$$
$$
  i   \hat{b}_{-}   \Phi_{1}
 - i  \hat{a}_{+}  \Phi_{3}     = M  \;   H_{2} \; .
\eqno(2.10)
$$

\subsection*{3.  Nonrelativistic approximation}

Excluding  non-dynamical  variables
$\Phi_{0}, H_{1}, H_{2}, H_{3}$ with the help of equations
$$
 -\hat{b}_{-}   E_{1}   - \hat{a}_{+}    E_{3}      = M   \;  \Phi_{0} \; ,
$$
$$
-  i  \; \hat{a}   \Phi_{2} = M  \; H_{1}\;,
$$
$$
 i   \hat{b}_{-}   \Phi_{1}  - i  \hat{a}_{+}  \Phi_{3}     = M  \;   H_{2} \;
 ,$$
 $$
  i   \; \hat{b}   \Phi_{2}    =  M  \;    H_{3}\;,
  \eqno(3.1)
  $$

\noindent  we get 6 equations (grouping them in pairs)
    $$
        i \hat{a}  \;
        ( i   \hat{b}_{-}   \Phi_{1}  - i  \hat{a}_{+}  \Phi_{3})
         + i \epsilon  \;   M E_{1}
  = M^{2}     \Phi _{1}\;,
$$
$$
     \hat{a} \;
      (-\hat{b}_{-}   E_{1}   - \hat{a}_{+}    E_{3}
  - i  \epsilon  \;  M\Phi_{1} = M ^{2}     e_{1}\;,
\eqno(3.2a)
$$

$$
   - i  \hat{b}_{-}  \; ( -  i   \; \hat{a} \;  \Phi_{2} )    +  i \hat{a}_{+}
(i   \; \hat{b} \;  \Phi_{2})
    + i \epsilon  \;   M E_{2}   = M ^{2}   \Phi_{2}\;,
    $$
$$
-i \epsilon \;   M \Phi_{2}       = M^{2}    E_{2} \;,
\eqno(3.2b)
$$

$$
  - i \; \hat{b}\;
  ( i   \hat{b}_{-}  \Phi_{1}  - i  \hat{a}_{+}  \Phi_{3})
+ i \epsilon  \;   M E_{3}  = M^{2}     \Phi_{3}\; ,
$$
$$
    \hat{b} \;
     (-\hat{b}_{-}   E_{1}   - \hat{a}_{+}    E_{3} )   -i \epsilon  \; M \Phi _{3}  = M  ^{2}  E_{3}\;,
\eqno(3.2c)
$$

Now we introduce big and small constituents
$$
\Phi_{1} = \Psi_{1} + \psi_{1}\; , \qquad i E_{1} =  \Psi_{1} - \psi_{1}\; ,
$$
$$
\Phi_{2} = \Psi_{2} + \psi_{2}\; , \qquad i E_{2} =  \Psi_{2} - \psi_{2}\; ,
$$
$$
\Phi_{3} = \Psi_{3} + \psi_{3}\; , \qquad i E_{3} =  \Psi_{3} - \psi_{3}\; ;
$$

\noindent
besides we should separate the rest energy by formal change $\epsilon \Longrightarrow \epsilon + M $;
summing and subtracting equation within each pair  (3.2) and ignoring small
constituents
 $\psi_{i}$ we arrive at three equations for big  components
$$
\left ( - 2 \; \hat{a} \hat{b}_{-}      + 2\epsilon M  \right ) \Psi_{1}    = 0 \; ,
$$
$$
\left ( -   (\hat{b}_{-} \hat{a}  + \hat{a}_{+} \hat{b} )
+2\epsilon M     \right )  \Psi_{2}
  =0 \; ,
$$
$$
\left ( -  2  \hat{b} \hat{a}_{+}
  +2\epsilon M \right )  \Psi_{3}    = 0 \; .
  \eqno(3.4)
$$

\noindent
It is a needed Pauli-like system  for the spin 1 particle.

Explicitly they read
$$
\left[{d^{2}\over dr^{2}}+{\cosh r\over \sinh r}\,{d\over dr}-{1\over \sinh r}\,{d\nu\over dr}-{1-2\,\nu\,\cosh r\over \sinh^{2}r}-{\nu^{2}\over \sinh^{2}r}+2\,\epsilon\, M \right]\Psi_{1}=0\,,
$$
$$
\left[{d^{2}\over dr^{2}}+{\cosh r\over \sinh r}\,{d\over dr}-{\nu^{2}\over \sinh^{2}r}+2\,\epsilon \, M\right]\Psi_{2}=0\,,
$$
$$
\left[{d^{2}\over dr^{2}}+{\cosh r\over \sinh r}\,{d\over dr}
+{1\over \sinh r}\,{d\nu\over dr}-{1+2\,\nu\,\cosh r\over \sinh^{2}r}-
{\nu^{2}\over \sinh^{2}r}+2\,\epsilon\, M \right]\Psi_{3}=0\, .
$$
$$
\eqno(3.5)
$$

\noindent Allowing for  $\nu (r) =m+B\,(\cosh r-1)$ we arrive at
$$
\left[{d^{2}\over dr^{2}}+{\cosh r\over \sinh r}\,{d\over dr}-B-
{1-2\, [ m+B\,(\cosh r-1) ]\,\cosh r\over \sinh^{2}r}-
\right.
$$
$$
\left. - {[m+B\,(\cosh r-1)]^{2}\over \sinh^{2}r}+2\,\epsilon\, M \right]\Psi_{1}=0\,,
$$

$$
\left[{d^{2}\over dr^{2}}+{\cosh r\over \sinh r}\,{d\over dr}-
{[ m+B\,(\cosh r-1)]^{2}\over \sinh^{2}r}+2\,\epsilon \, M\right]\Psi_{2}=0\,,
$$

\vspace{3mm}

$$
\left[{d^{2}\over dr^{2}}+{\cosh r\over \sinh r}\,{d\over dr}+B-{1+2\,[m +B\,(\cosh r-1)] \,\cosh r\over \sinh^{2}r}-
\right.
$$
$$
\left. -
{[m+B\,(\cosh r-1)] ^{2}\over \sinh^{2}r}+2\,\epsilon\, M \right]\Psi_{3}=0\,.
\eqno(3.6)
$$

The first and the third equations are symmetric with respect to  formal change
$
m \Longrightarrow -m\,, \;B \Longrightarrow - B$.

In the new variable
$
1-\cosh r =2 \,y
$, they look
$$
y\,(1-y)\,{d^{2}\Psi_{1}\over dy^{2}}+(1-2\,y)\,{dB_{1}\over dy}+
$$
$$
+\left[B^{2}-B-2\,\epsilon\,M-{1\over 4}\,{(2\,B-m-1)^{2}\over1-y}-{1\over 4}\,{(m-1)^{2}\over y}\right]\Psi_{1}=0\,,
$$
$$
\eqno(3.7a)
$$

$$
y\,(1-y)\,{d^{2}\Psi_{2}\over dy^{2}}+(1-2\,y)\,{dB_{2}\over dy}+
$$
$$
+\left[B^{2}-2\,\epsilon\,M-{1\over 4}\,{(2\,B-m)^{2}\over1-y}-{1\over 4}\,{m^{2}\over y}\right]\Psi_{2}=0\,,
$$
$$
\eqno(3.7b)
$$

$$
y\,(1-y)\,{d^{2}\Psi_{3}\over dy^{2}}+(1-2\,y)\,{dB_{3}\over dy}+
$$
$$
+\left[B^{2}+B-2\,\epsilon\,M-{1\over 4}\,{(2\,B-m+1)^{2}\over1-y}-{1\over 4}\,{(m+1)^{2}\over y}\right]\Psi_{3}=0\,.
$$
$$
\eqno(3.7c)
$$

Eq.  $(3.7a)$ with  the substitution $$
\Psi_{1} = y^{C_{1}} (1-y) ^{A_{1}}  f_{1}
$$

\noindent leads to
$$
y\,(1-y)\,{d^{2}\Psi_{1}\over dy^{2}}+[2\,C_{1}+1-(2\,A_{1}+2\,C_{1}+2)\,y]\,{dB_{1}\over dy}+
$$
$$
+\left[B^{2}-B-2\,\epsilon\,M-(A_{1}+C_{1})\,(A_{1}+C_{1}+1)+\right.
$$
$$\left.+\,
{1\over 4}\,{4\,A_{1}^{2}-(2\,B-m-1)^{2}\over1-y}+{1\over 4}\,{4\,C_{1}^{2}-(m-1)^{2}\over y}\right]\Psi_{1}=0\,.
\eqno(3.8)
$$

\noindent At  $A_{1},\,C_{1}$ obeying
$$
A_{1}=\pm\,{1\over 2}\,(2\,B-m-1)\,,\qquad C_{1}= \pm\,{1\over 2}\,(m-1)\,,
$$

\noindent eq.  $(3.8)$ becomes simpler
$$
y\,(1-y)\,{d^{2}\Psi_{1}\over dy^{2}}+[2\,C_{1}+1-(2\,A_{1}+2\,C_{1}+2)\,y]\,{dB_{1}\over dy}+
$$
$$
+\left[B^{2}-B-2\,\epsilon\,M-(A_{1}+C_{1})\,(A_{1}+C_{1}+1)\right]\Psi_{1}=0 \; ,
\eqno(3.9a)
$$

\noindent what is hypergeometric equation with parameters
$$
\alpha_{1}=A_{1}+C_{1}+{1\over 2}+\sqrt{B^{2}-B-2\,\epsilon\,M+{1\over 4}}\,,
$$
$$
\beta_{1}=A_{1}+C_{1}+{1\over 2}-\sqrt{B^{2}-B-2\,\epsilon\,M+{1\over 4}}\,,
$$
$$
\gamma_{1}=2\,C_{1}+1\,.
\eqno(3.9b)
$$

To have finite and single-valued solutions
one must impose restrictions  $A_{1}<0$, $C_{1}>0$. Besides,
one must get $n$-order polynomials
and satisfy the inequality  $A_{1}+C_{1}+n<0$.

Four different possibilities  for $A_{1},\,C_{1}$ are (for definiteness let it be
 $B>0$):
$$
1.\qquad  A_{1}\,=\,- {1\over 2}\,(2\,B-m-1) \,,\qquad C_{1}\,=\,- {1\over 2}\,(m-1) \,,
$$
$$
2.\qquad  A_{1}\,=\, +{1\over 2}\,(2\,B-m-1) \,,\qquad C_{1}\,=\,-{1\over 2}\,(m-1) \,,
$$
$$
3.\qquad  A_{1}\,=\,+ {1\over 2}\,(2\,B-m-1) \,,\qquad C_{1}\,=\, +{1\over 2}\,(m-1) \,,
$$
$$
4.\qquad  A_{1}\,=\, -{1\over 2}\,(2\,B-m-1) \,,\qquad C_{1}\,=\, +{1\over 2}\,(m-1) \,.
$$

To describe bound state, only  variants 1 and  4 are appropriate:
$$
\mbox{ 1\; ,}\qquad m\,<0\,,
$$
$$
\alpha_{1}=-B+{3\over 2}+\sqrt{B^{2}-B-2\,\epsilon\,M+{1\over 4}}\,,
$$
$$
\beta_{1}=-B+{3\over 2}-\sqrt{B^{2}-B-2\,\epsilon \,M+{1\over 4}}\,,
$$
$$
\gamma_{1}=-m+2\,,
$$
$$
\mbox{spectrum}\qquad \alpha_{1}=-n\,,\qquad \sqrt{B^{2}-B-2\,\epsilon\,M+{1\over 4}}=B-{3\over 2}-n\,,
\eqno(3.10a)
$$
$$
\epsilon\,M
=B-1+n\,\left(B-{3\over 2}-{n\over 2}\right)\,;$$

\vspace{5mm}

$$\mbox{ 4\; ,}\qquad 0<m<B\,,$$
$$
\alpha_{1}=-B+m+{1\over 2}+\sqrt{B^{2}-B-2\,\epsilon\,M+{1\over 4}}\,,
$$
$$
\beta_{1}=-B+m+{1\over 2}-\sqrt{B^{2}-B-2\,\epsilon\,M+{1\over 4}}\,,
$$
$$
\gamma_{1}=m\,,
$$
$$
\mbox{spectrum}\qquad \alpha_{1}=-n\,,\qquad \sqrt{B^{2}-B-2\,\epsilon\,M+{1\over 4}}=B-{1\over 2}-(n+m)\,,
\eqno(3.10b)
$$
$$
\epsilon\,M=(m+n)\,\left(B-{1\over 2}-{1\over2}\,(m+n)\right)\,.
$$

Formulas  (3.10a,b)  can be jointed into single one
$$
\sqrt{B^{2}-B - 2\,\epsilon\,M+{1\over 4}} = -n - {1 \over 2} -  { \mid 2B - m -1  \mid +  \mid m-1 \mid \over 2} \; .
\eqno(3.10c)
$$

\vspace{5mm}

From eq. $(3.7b)$ with the substitution $$
\Psi_{2} = y^{C_{2}} (1-y) ^{A_{2}}  f_{2}
$$

\noindent we get
$$
y\,(1-y)\,{d^{2}f_{2}\over dy^{2}}+[2\,C_{2}+1-(2\,A_{2}+2\,C_{2}+2)\,y]\,{df_{2}\over dy}+
$$
$$
+\left[B^{2}-2\,\epsilon\,M-(A_{2}+C_{2})\,(A_{2}+C_{2}+1)+\right.
$$
$$\left.+\,
{1\over 4}\,{4\,A_{2}^{2}-(2\,B-m)^{2}\over1-y}+{1\over 4}\,{4\,C_{2}^{2}-m^{2}\over y}\right]f_{2}=0\,.
\eqno(3.11)
$$

\noindent At
$$
A_{2}=\pm\,{1\over 2}\,(2\,B-m)\,,\qquad C_{2}= \pm\,{m\over 2}\,,
$$

\noindent eq.  $(3.11)$ becomes simpler
$$
y\,(1-y)\,{d^{2}f_{2}\over dy^{2}}+[2\,C_{2}+1-(2\,A_{2}+2\,C_{2}+2)\,y]\,{df_{2}\over dy}+
$$
$$
+\left[B^{2}-2\,\epsilon\,M-(A_{2}+C_{2})\,(A_{2}+C_{2}+1)\right]f_{2}=0
\eqno(3.12a)
$$

\noindent which is  recognized as of hypergeometric type
$$
\alpha_{2}=A_{2}+C_{2}+{1\over 2}+\sqrt{B^{2}-2\,\epsilon\,M+{1\over 4}}\,,
$$
$$
\beta_{2}=A_{2}+C_{2}+{1\over 2}-\sqrt{B^{2}-2\,\epsilon\,M+{1\over 4}}\,,
$$
$$
\gamma_{2}=2\,C_{2}+1\,.
\eqno(3.12b)
$$

From four variants
$$
1.\qquad  A_{2}\,=\,- {1\over 2}\,(2\,B-m) \,,\qquad C_{2}\,=\,- {m\over 2} \,,
$$
$$
2.\qquad  A_{2}\,=\, +{1\over 2}\,(2\,B-m) \,,\qquad C_{2}\,=\,-{m\over 2} \,,
$$
$$
3.\qquad  A_{2}\,=\,+ {1\over 2}\,(2\,B-m) \,,\qquad C_{2}\,=\, +{m\over 2} \,,
$$
$$
4.\qquad  A_{2}\,=\, -{1\over 2}\,(2\,B-m) \,,\qquad C_{2}\,=\, +{m\over 2} \,
$$

\noindent only  1 and  4 seem to be approptiate to describe bound states:
$$
\mbox{ 1\; ,}\qquad m\,<0\,,
$$
$$
\alpha_{2}=-B+{1\over 2}+\sqrt{B^{2}-2\,\epsilon\,M+{1\over 4}}\,,
$$
$$
\beta_{2}=-B+{1\over 2}-\sqrt{B^{2}-2\,\epsilon \,M+{1\over 4}}\,,
$$
$$
\gamma_{2}=-m+1\,,
$$
$$
\mbox{spectrum}\qquad \alpha_{2}=-n\,,\qquad \sqrt{B^{2}-2\,\epsilon\,M+{1\over 4}}=B-{1\over 2}-n\,,
\eqno(3.13a)
$$
$$
\epsilon\,M
={B\over2}+n\,\left(B-{1\over 2}-{n\over 2}\right)\,;$$

\vspace{5mm}

$$\mbox{ 4\; ,}\qquad 0<m<B\,,$$
$$
\alpha_{2}=-B+m+{1\over 2}+\sqrt{B^{2}-2\,\epsilon\,M+{1\over 4}}\,,
$$
$$
\beta_{2}=-B+m+{1\over 2}-\sqrt{B^{2}-2\,\epsilon\,M+{1\over 4}}\,,
$$
$$
\gamma_{2}=m+1\,,
$$
$$
\mbox{spectrum}\qquad \alpha_{2}=-n\,,\qquad \sqrt{B^{2}-2\,\epsilon\,M+{1\over 4}}=B-{1\over 2}-(n+m)\,,
\eqno(3.13b)
$$
$$
\epsilon\,M={B\over 2}+(m+n)\,\left(B-{1\over 2}-{1\over2}\,(m+n)\right)\,.
$$

Formulas   (3.13a,b)  can be joint into a single one
$$
\sqrt{B^{2} - 2\,\epsilon\,M+{1\over 4}} = -n - {1 \over 2} - { \mid 2B - m   \mid +  \mid m\mid \over 2} \; .
\eqno(3.13c)
$$

The region for allowed values of  $m$ for bound states can be
illustrated by
Fig. 1.

\vspace{-20mm}

\unitlength=0.6mm
\begin{picture}(160,100)(-90,0)

\special{em:linewidth 0.4pt} \linethickness{0.4pt}

\put(-70,0){\vector(+1,0){140}}  \put(+70,-5){$m$}
\put(0,-50){\vector(0,+1){100}}   \put(+2,+53){$\mid m \mid -\mid 2B -m \mid <0$}

\put(0,0){\line(+1,+1){40}}
\put(0,0){\line(-1,+1){40}}
\put(+45,+35){$\mid m \mid $}

\put(+30,0){\circle*{2}}  \put(+35,+2){$2B$}

\put(+15,0){\circle*{2}}   \put(+10,+2){$B$}


\put(+30,0){\line(-1,-1){70}}  \put(+30,0){\line(+1,-1){30}}  \put(-60,-50){$-\mid 2B -m \mid $}

\put(+15,+0.3){\line(-1,0){85}}
\put(+15,+0.2){\line(-1,0){85}}
\put(+15,+0.1){\line(-1,0){85}}
\put(+15,-0.1){\line(-1,0){85}}
\put(+15,-0.2){\line(-1,0){85}}
\put(+15,-0.3){\line(-1,0){85}}

\end{picture}

\vspace{55mm}

\begin{center}
{\bf Fig. 1.  Bound states at  $B>0: \; m  < B $ }
\end{center}

At  $B<0$, we shoul have different Fig. . 2.

\vspace{-20mm}

\unitlength=0.7mm
\begin{picture}(160,100)(-90,0)

\special{em:linewidth 0.4pt} \linethickness{0.4pt}

\put(-70,0){\vector(+1,0){140}}  \put(+70,-5){$m$}
\put(0,-50){\vector(0,+1){100}}   \put(+2,+53){$\mid m \mid -\mid 2B -m \mid <0$}

\put(0,0){\line(+1,+1){40}}
\put(0,0){\line(-1,+1){40}}
\put(+45,+35){$\mid m \mid $}

\put(-30,0){\circle*{2}}  \put(-35,+2){$2B$}

\put(-15,0){\circle*{2}}   \put(-15,+2){$B$}


\put(-30,0){\line(+1,-1){70}}  \put(-30,0){\line(-1,-1){30}}  \put(-75,-40){$-\mid 2B -m \mid $}

\put(-15,+0.2){\line(+1,0){85}}
\put(-15,+0.1){\line(+1,0){85}}
\put(-15,-0.1){\line(+1,0){85}}
\put(-15,-0.2){\line(+1,0){85}}

\end{picture}

\vspace{50mm}

\begin{center}
{\bf Fig.  2.  Bound states at  $B<0: \; B < m $ }
\end{center}

\vspace{5mm}

Similar Figures can be given in connection with the functions $\Psi_{1}(y)$ and $\Psi_{3}$ as well.

In case of $(3.7c)$, with substitution
$$
\Psi_{3} = y^{C_{3}} (1-y) ^{A_{3}}  f_{3}\,,
$$

\noindent we will obtain
$$
y\,(1-y)\,{d^{2}f_{3}\over dy^{2}}+[2\,C_{3}+1-(2\,A_{3}+2\,C_{3}+2)\,y]\,{df_{3}\over dy}+
$$
$$
+\left[B^{2}+B-2\,\epsilon\,M-(A_{3}+C_{3})\,(A_{3}+C_{3}+1)+\right.
$$
$$\left.+\,
{1\over 4}\,{4\,A_{3}^{2}-(2\,B-m+1)^{2}\over1-y}+{1\over 4}\,{4\,C_{3}^{2}-(m+1)^{2}\over y}\right]f_{3}=0\,.
\eqno(3.14)
$$

At $A_{3},\,C_{3}$
$$
A_{3}=\pm\,{1\over 2}\,(2\,B-m+1)\,,\qquad C_{3}= \pm\,{1\over 2}\,(m+1)\,,
$$

\noindent eq. $(3.14)$ will read
$$
y\,(1-y)\,{d^{2}f_{3}\over dy^{2}}+[2\,C_{3}+1-(2\,A_{3}+2\,C_{3}+2)\,y]\,{df_{3}\over dy}+
$$
$$
+\left[B^{2}+B-2\,\epsilon\,M-(A_{3}+C_{3})\,(A_{3}+C_{3}+1)\right]f_{3}=0
\eqno(3.15a)
$$

\noindent that is a hypergeometric equation
$$
\alpha_{3}=A_{3}+C_{3}+{1\over 2}+\sqrt{B^{2}+B-2\,\epsilon\,M+{1\over 4}}\,,
$$
$$
\beta_{3}=A_{3}+C_{3}+{1\over 2}-\sqrt{B^{2}+B-2\,\epsilon\,M+{1\over 4}}\,,
$$
$$
\gamma_{3}=2\,C_{3}+1\,.
\eqno(3.15b)
$$

From four possibilities
$$
1.\qquad  A_{3}\,=\,- {1\over 2}\,(2\,B-m+1) \,,\qquad C_{3}\,=\,- {1\over 2}\,(m+1) \,,
$$
$$
2.\qquad  A_{3}\,=\, +{1\over 2}\,(2\,B-m+1) \,,\qquad C_{3}\,=\,-{1\over 2}\,(m+1) \,,
$$
$$
3.\qquad  A_{3}\,=\,+ {1\over 2}\,(2\,B-m+1) \,,\qquad C_{3}\,=\, +{1\over 2}\,(m+1) \,,
$$
$$
4.\qquad  A_{3}\,=\, -{1\over 2}\,(2\,B-m+1) \,,\qquad C_{3}\,=\, +{1\over 2}\,(m+1) \,.
$$

only 1  and  4 are appropriate to describe bound states:
$$
\mbox{ 1\; ,}\qquad m\,<0\,,
$$
$$
\alpha_{3}=-B-{1\over 2}+\sqrt{B^{2}+B-2\,\epsilon\,M+{1\over 4}}\,,
$$
$$
\beta_{3}=-B-{1\over 2}-\sqrt{B^{2}+B-2\,\epsilon \,M+{1\over 4}}\,,
$$
$$
\gamma_{3}=-m\,,
$$
$$
\mbox{spectrum}\qquad \alpha_{3}=-n\,,\qquad \sqrt{B^{2}+B-2\,\epsilon\,M+{1\over 4}}=B+{1\over 2}-n\,,
\eqno(3.16a)
$$
$$
\epsilon\,M
=n\,\left(B+{1\over 2}-{n\over 2}\right)\,;$$

\vspace{5mm}

$$\mbox{ 4\; ,}\qquad 0<m<B\,,$$
$$
\alpha_{3}=-B+m+{1\over 2}+\sqrt{B^{2}+B-2\,\epsilon\,M+{1\over 4}}\,,
$$
$$
\beta_{3}=-B+m+{1\over 2}-\sqrt{B^{2}+B-2\,\epsilon\,M+{1\over 4}}\,,
$$
$$
\gamma_{3}=m+2\,,
$$
$$
\mbox{spectrum}\qquad \alpha_{3}=-n\,,\qquad \sqrt{B^{2}+B-2\,\epsilon\,M+{1\over 4}}=B-{1\over 2}-(n+m)\,,
\eqno(3.16b)
$$
$$
\epsilon\,M=B+(m+n)\,\left(B-{1\over 2}-{1\over2}\,(m+n)\right)\,.
$$

Again, formulas (3.16a,b)  can be joint into a single one
$$
\sqrt{B^{2}+ B - 2\,\epsilon\,M+{1\over 4}} = -n - {1 \over 2} - { \mid 2B - m +1  \mid +  \mid m+1 \mid \over 2} \; .
\eqno(3.16c)
$$

\subsection*{4. Solution of radial equations in relativistic case
}

Let start with eqs.  (2.4)--(2.5)
$$
  -\hat{b}_{-}   E_{1}   - \hat{a}_{+}    E_{3}
 = M   \; \Phi_{0} \; ,
$$
$$
   - i  \hat{b}_{-}    H_{1}    +  i \hat{a}_{+} \; H_{3}
    + i \epsilon  \;   E_{2}   = M  \;  \Phi_{2}\;,
    $$
    $$
        i  \hat{a}     H_{2} +i \epsilon  \;   E_{1}
  = M  \;   \Phi _{1}\;,
$$
$$
  - i  \hat{b}  H_{2} +i \epsilon  \;   E_{3}
= M    \; \Phi_{3}\; ,
\eqno(4.1)
$$
$$
    \hat{a}  \Phi_{0}  - i  \epsilon  \;  \Phi_{1} = M \;     E_{1}\;,
$$
$$
-  i  \hat{a}  \Phi_{2}  = M  \; H_{1}\;,
$$
$$
   \hat{b}
\Phi_{0}  -i \epsilon  \;\Phi _{3}  = M    \; E_{3}\;,
$$
$$
  i   \hat{b}   \Phi_{2}
   =  M  \;    H_{3}\;.
$$
$$
-i \epsilon   \Phi_{2}    = M  \;  E_{2} \;,
$$
$$
  i   \hat{b}_{-}   \Phi_{1}
 - i  \hat{a}_{+}  \Phi_{3}     = M  \;   H_{2} \; ,
\eqno(4.2)
$$

 Excluding six components  $E_{i}, H_{i}$, we derive
four second order equations for  $\Phi _{a}$:
$$
(- \hat{b}_{-} \hat{a} - \hat{a}_{+} \hat{b}  + \epsilon^{2} - M^{2} ) \Phi_{2} = 0 \; ,
$$
$$
(- \hat{b}_{-} \hat{a} - \hat{a}_{+} \hat{b}
 - M^{2} ) \Phi_{0} + i \epsilon ( \hat{b}_{-} \Phi_{1} + \hat{a}_{+}\Phi_{3})=0\; ,
$$
$$
(- \hat{a} \hat{b}_{-} + \epsilon^{2} - M^{2}) \Phi_{1} +
\hat{a} \hat{a}_{+} \Phi_{3} + i \epsilon \hat{a} \Phi_{0} = 0 \; ,
$$
$$
(- \hat{b} \hat{a}_{+} + \epsilon^{2} - M^{2}) \Phi_{3} +
\hat{b} \hat{b}_{-} \Phi_{1} + i \epsilon \hat{b} \Phi_{0} = 0 \; .
\eqno(4.3)
$$

Once, it should be noted existence of a simple solution of the system
$$
\Phi_{0} = 0\;, \qquad \Phi_{1}=0\;, \qquad \Phi_{3}=0 \; ,
$$
$$
(- \hat{b}_{-} \hat{a} - \hat{a}_{+} \hat{b}  + \epsilon^{2} - M^{2} ) \Phi_{2} = 0 \; .
\eqno(4.4a)
$$

\noindent and simple expressions for tensors components
$$
E_{1} = 0 \;, \qquad     H_{1} = -  i M^{-1}  \hat{a} \; \Phi_{2}  \;,
$$
$$
E_{3} =0 \; , \qquad         H_{3}= i  M^{-1} \hat{b}  \; \Phi_{2}
    \; ,
   $$
   $$
 E_{2} = -i \epsilon  M^{-1}  \Phi_{2}      \; , \qquad H_{2} = 0 \; .
\eqno(4.4b)
$$

Lets us turn to (4.3) and act on  the third equation from the left by operator $\hat{b}_{-}$,
and on the forth  equation by operator $\hat{a}_{+}$. Thus, introducing the notation
$$
\hat{b}_{-} \Phi_{1}  = Z_{1} \; , \qquad \hat{a}_{+} \Phi_{3} = Z_{3} \;,
$$

\noindent instead of (4.3)  we obtain
$$
(- \hat{b}_{-} \hat{a} - \hat{a}_{+} \hat{b}  + \epsilon^{2} - M^{2} ) \Phi_{2} = 0 \; ,
$$
$$
(- \hat{b}_{-} \hat{a} - \hat{a}_{+} \hat{b}
 - M^{2} ) \Phi_{0} + i \epsilon ( Z_{1} + Z_{3} )=0\; ,
$$
$$
(- \hat{b}_{-} \hat{a}  + \epsilon^{2} - M^{2})  Z_{1}  +
\hat{b}_{-} \hat{a} Z_{3} + i \epsilon \hat{b}_{-} \hat{a} \Phi_{0} = 0 \; ,
$$
$$
(-  \hat{a}_{+} \hat{b} + \epsilon^{2} - M^{2}) Z_{3}  +
\hat{a}_{+} \hat{b} Z_{1}  + i \epsilon  \hat{a}_{+} \hat{b} \Phi_{0} = 0 \; .
\eqno(4.5)
$$

\noindent
Instead of $Z_{1}, Z_{3}$, let us introduce new  functions
$$
Z_{1} ={f + g \over  2} \; , \qquad  Z_{3} ={f - g \over  2} \; ,
$$
$$Z_{1} +Z_{3} = f \;, \qquad Z_{1} - Z_{3} = g \; ;
$$

\noindent
the the above  system reads
$$
(- \hat{b}_{-} \hat{a} - \hat{a}_{+} \hat{b}  + \epsilon^{2} - M^{2} ) \Phi_{2} = 0 \; ,
$$
$$
(- \hat{b}_{-} \hat{a} - \hat{a}_{+} \hat{b}
 - M^{2} ) \Phi_{0} + i \epsilon \; f =0\; ,
$$
$$
- \hat{b}_{-} \hat{a}  {f + g \over  2}   + (\epsilon^{2} - M^{2})  {f + g \over  2}    +
\hat{b}_{-} \hat{a} {f - g \over  2}  + i \epsilon \hat{b}_{-} \hat{a} \Phi_{0} = 0 \; ,
$$
$$
-  \hat{a}_{+} \hat{b}  {f - g \over  2} + (\epsilon^{2} - M^{2}) {f - g \over  2}   +
\hat{a}_{+} \hat{b} {f + g \over  2}  + i \epsilon  \hat{a}_{+} \hat{b} \Phi_{0} = 0 \; .
\eqno(4.6)
$$

\noindent
After elementary manipulations with  equation 3 and  4  we get
$$
(- \hat{b}_{-} \hat{a} - \hat{a}_{+} \hat{b}  + \epsilon^{2} - M^{2} ) \Phi_{2} = 0 \; ,
$$
$$
(- \hat{b}_{-} \hat{a} - \hat{a}_{+} \hat{b}
 - M^{2} ) \Phi_{0} + i \epsilon \; f =0\; ,
$$
$$
- \hat{b}_{-} \hat{a}   g    + (\epsilon^{2} - M^{2})  {f + g \over  2}      + i \epsilon \hat{b}_{-} \hat{a} \Phi_{0} = 0 \; ,
$$
$$
  \hat{a}_{+} \hat{b}   g  + (\epsilon^{2} - M^{2}) {f - g \over  2}
    + i \epsilon  \hat{a}_{+} \hat{b} \Phi_{0} = 0 \; .
$$

Now, summing and subtracting  equations 3 and 4, we obtain
$$
(- \hat{b}_{-} \hat{a} - \hat{a}_{+} \hat{b}  + \epsilon^{2} - M^{2} ) \Phi_{2} = 0 \; ,
$$
$$
(- \hat{b}_{-} \hat{a} - \hat{a}_{+} \hat{b}
 - M^{2} ) \Phi_{0} + i \epsilon \; f =0\; ,
$$
$$
( - \hat{b}_{-} \hat{a}   + \hat{a}_{+} \hat{b})  g    + (\epsilon^{2} - M^{2})  f
      + i \epsilon  ( \hat{b}_{-} \hat{a} + \hat{a}_{+} \hat{b} )\;  \Phi_{0} = 0 \; ,
$$
$$
( - \hat{b}_{-} \hat{a} - \hat{a}_{+} \hat{b})  g    + (\epsilon^{2} - M^{2})  g
    + i \epsilon ( \hat{b}_{-} \hat{a}- \hat{a}_{+} \hat{b})  \Phi_{0} = 0 \; ,
\eqno(4.7)
$$

\noindent
Taking into account identities
$$
- \hat{b}_{-} \hat{a} - \hat{a}_{+} \hat{b}= \Delta_{2} = ...
$$
$$
- \hat{b}_{-} \hat{a}   + \hat{a}_{+} \hat{b}=  2B
\eqno(4.8)
$$

\noindent
we arrive at the system
$$
(\Delta_{2}  + \epsilon^{2} - M^{2} ) \; \Phi_{2} = 0 \; ,
\eqno(4.9)
$$

$$
( \Delta_{2}  - M^{2} ) \; \Phi_{0} + i \epsilon \; f =0\; ,
$$
$$
2B  \;  g    + (\epsilon^{2} - M^{2})  f
      -  i \epsilon  \Delta_{2} \;  \Phi_{0} = 0 \; ,
$$
$$
\Delta_{2}  g    + (\epsilon^{2} - M^{2})  g
    - 2i \epsilon B \;   \Phi_{0} = 0 \; ,
\eqno(4.10)
$$

From the second equation, with the use of expression for $\Delta_{2} \Phi_{0}$ according to the first equation,
we  derive linear relation between three functions
$$
2B\; g -M^{2} f - i\epsilon M^{2} \Phi_{0} = 0 \; .
\eqno(4.11)
$$

\noindent Let us exclude  $f$
$$
 f = {2B \over M^{2}} \; g  - i\epsilon  \Phi_{0}
$$

\noindent so we get
$$
(\Delta_{2}     + \epsilon^{2} - M^{2}) \;  g
    = 2i \epsilon B \;   \Phi_{0} \; ,
    $$
    $$
(\Delta_{2}     + \epsilon^{2} - M^{2}) \; \Phi_{0}  = - {2i\epsilon B  \over  M^{2}} \; g  .
\eqno(4.12)
$$

\noindent
With notation  $\gamma = \epsilon^{2} / M^{2}$, the system can be presented in a matrix form as follows
$$
(\Delta_{2}     + \epsilon^{2} - M^{2})
\left | \begin{array}{r}
g  \\
\epsilon \; \Phi_{0}
\end{array} \right | =
\left | \begin{array}{rr}
0 & 2iB  \\
-2iB \gamma  & 0
\end{array} \right |
\left | \begin{array}{r}
g  \\
\epsilon \; \Phi_{0}
\end{array} \right | .
\eqno(4.13)
$$

\noindent
or symbolically
$$
\Delta f = A f \qquad  \Delta f ' = S AS^{-1} \;f'\;, \qquad f ' = S f \; .
$$

It remains to find a transformation reducing the matrix $A$ to a diagonal form
$$
S AS^{-1} = \left | \begin{array}{cc}
\lambda_{1} & 0 \\
0 & \lambda_{2}
\end{array} \right | , \qquad S = \left | \begin{array}{cc}
a & d \\
c & b
\end{array} \right | \; ;
$$

\noindent  the problem is equivalent to the linear system
$$
-\lambda_{1}  \; a - 2i\gamma B \; d  = 0 \; ,
$$
$$
2iB \; a  - \lambda_{1} \; d = 0 \; ;
$$

$$
-\lambda_{2}  \; c - 2i\gamma B \; b  = 0 \; ,
$$
$$
2iB \; c  - \lambda_{2} \; b = 0 \; .
$$

Its solutions can be chosen in the form
$$
  \lambda_{1} =  + {2\epsilon B \over M} \;, \qquad \lambda_{2} =
-  {2\epsilon B \over M} \;,
$$
$$
S= \left | \begin{array}{cc}
\epsilon & +iM \\
\epsilon & -iM
\end{array} \right |, \qquad
S^{-1}  = {1 \over -2i\epsilon M}
\left | \begin{array}{cc}
-i M & -i M \\
-\epsilon & \epsilon
\end{array} \right | .
\eqno(4.14)
$$

New (primed) function satisfy the following equations

$$
1) \qquad \left (\Delta_{2}     + \epsilon^{2} - M^{2} - {2\epsilon B \over M} \right  ) \; g'  = 0 \; ,
\eqno(4.15a)
$$
$$
2) \qquad \left (\Delta_{2}     + \epsilon^{2} - M^{2} + {2\epsilon B \over M}  \right  ) \;  \Phi_{0}'  = 0   \; .
\eqno(4.15b)
$$

they are independent from each other,  therefore there exist two
solutions
$$
1) \qquad  g' \neq 0 , \qquad \Phi'_{0}= 0 \; ,
\eqno(4.16a)
$$
$$
2) \qquad  g' =0  , \qquad \Phi'_{0} \neq  0 \; .
\eqno(4.16b)
$$

The initial functions for these two cases assume respectively the form
$$
g = {1 \over 2\epsilon } g' + {1 \over 2i\epsilon} \epsilon \Phi'_{0} \; ,\qquad
\epsilon \Phi_{0} = {1 \over  2i M} g' - {1  \over 2iM} \epsilon\Phi_{0}' \; .
\eqno(4.17)
$$

In cases 1) and  2) they assume respectively the form

$$
1)  \qquad  g = {1 \over 2\epsilon } g'  \; ,\qquad
\epsilon \Phi_{0} = {1 \over  2i M} g' \; .
\eqno(4.18a)
$$

$$
2) \qquad  g =  {1 \over 2i\epsilon} \epsilon \Phi'_{0} \; ,\qquad
\epsilon \Phi_{0} =  - {1  \over 2iM} \epsilon\Phi_{0}' \; .
\eqno(4.18b)
$$

To obtain explicit solutions for these differential equation, we need not any additional calculations,
instead it suffices to perform simple formal  changes as pointed below

$$
\left[{d^{2}\over dr^{2}}+{\cosh r\over \sinh r}\,{d\over dr}-
{[ m+B\,(\cosh r-1)]^{2}\over \sinh^{2}r}+2\,\epsilon \, M\right] f (r)  =0\,,
$$
$$
\sqrt{B^{2}-2\,\epsilon\,M+{1\over 4}} =- n - {1 \over 2} -  { \mid 2 B - m  \mid +  \mid m \mid \over 2}
\eqno(4.19)
$$

$$
2\,\epsilon \, M \qquad \Longrightarrow \qquad
\left \{ \begin{array}{l}
(\epsilon^{2} - M^{2}  - {2\epsilon B \over M} )\qquad  ---\;\;\;  \qquad (4.9)  \\
(\epsilon^{2} - M^{2} )\qquad  \qquad ---  \qquad \;\; (4.15a) \\
(\epsilon^{2} - M^{2} + {2\epsilon B \over M} )  \qquad ---  \qquad (4.15b)
\end{array} \right.
\eqno(4.20)
$$


\section{ Acknowledgment }

 Authors are  grateful  to  Dr. V.M. Red'kov   for stimulating discussion and  advices.

This  work was   supported   by the Fund for Basic Researches of Belarus
 F11M-152.

\end{document}